\def\year{2022}\relax
\documentclass[letterpaper]{article} 
\usepackage{aaai22}  
\usepackage{times}  
\usepackage{helvet}  
\usepackage{courier}  
\usepackage[hyphens]{url}  
\usepackage{graphicx} 
\usepackage{natbib}  
\usepackage{caption} 
\DeclareCaptionStyle{ruled}{labelfont=normalfont,labelsep=colon,strut=off} 
\usepackage{algorithm}
\usepackage{algorithmic}

\usepackage{newfloat}
\usepackage{listings}

\usepackage{amsmath}
\usepackage{multirow}
\usepackage{makecell}
\usepackage{tabularx}
\usepackage{subfigure}
\usepackage{amssymb}
\usepackage{color,xcolor}
\usepackage{booktabs}
\def\eg{\emph{e.g.}}
\def\ie{\emph{i.e.}}
\def\etc{\emph{etc}}

\floatstyle{ruled}
\newfloat{listing}{tb}{lst}{}
\floatname{listing}{Listing}
\title{Phonemic Adversarial Attack against Audio Recognition in Real World}
\author{
    Jiakai Wang\textsuperscript{\rm 1},
    Zhendong Chen\textsuperscript{\rm 2},
    Zixin Yin\textsuperscript{\rm 3},
    Qinghong Yang\textsuperscript{\rm 2,\thanks{Co-corresponding author}},
    Xianglong Liu\textsuperscript{\rm 1,3,$^\dagger$}
}
\affiliations{
    
    \textsuperscript{\rm 1} Zhongguancun Laboratory\\
    \textsuperscript{\rm 2} College of Software, Beihang University\\
    \textsuperscript{\rm 3} State Key Lab of Software Development Environment, Beihang University\\
    
}

\usepackage{bibentry}

\begin{document}

\maketitle

\begin{abstract}

Recently, adversarial attacks for audio recognition have attracted much attention. However, most of the existing studies mainly rely on the coarse-grain audio features at the instance level to generate adversarial noises, which leads to expensive generation time costs and weak universal attacking ability. Motivated by the observations that all audio speech consists of fundamental phonemes, this paper proposes a \textbf{p}honemic \textbf{a}dversarial a\textbf{t}tack (PAT) paradigm, which attacks the fine-grain audio features at the phoneme level commonly shared across audio instances, to generate phonemic adversarial noises, enjoying the more general attacking ability with fast generation speed. Specifically, for \textbf{accelerating the generation}, a phoneme density balanced sampling strategy is introduced to sample quantity less but phonemic features abundant audio instances as the training data via estimating the phoneme density, which substantially alleviates the heavy dependency on the large training dataset. Moreover, for \textbf{promoting universal attacking ability}, the phonemic noise is optimized in an asynchronous way with a sliding window, which enhances the phoneme diversity and thus well captures the critical fundamental phonemic patterns. By conducting extensive experiments, we comprehensively investigate the proposed PAT framework and demonstrate that it outperforms the SOTA baselines by large margins (\ie, at least \textbf{11}$\times$ speed up and \textbf{78\%} attacking ability improvement)\footnote{\textcolor{pink}{\url{https://anonymous.4open.science/r/Phonemic-Adversarial-Attack-1135}}}. 

\end{abstract}

\section{Introduction}

The bloom of deep learning techniques promotes the development of speech applications, \ie, automatic speech recognition (ASR), voiceprint recognition (VPR), \etc. However, these significant techniques are utilized for social bad, \eg, audio information stealing, therefore emerging requirements for protecting human privacy in recent years. Fortunately, adversarial attack, which is meticulously designed for misleading deep models, provides a feasible approach to address this urgent problem and attracts much attention of researchers.

\begin{figure}[!t]
\centering
\includegraphics[width=0.96\linewidth]{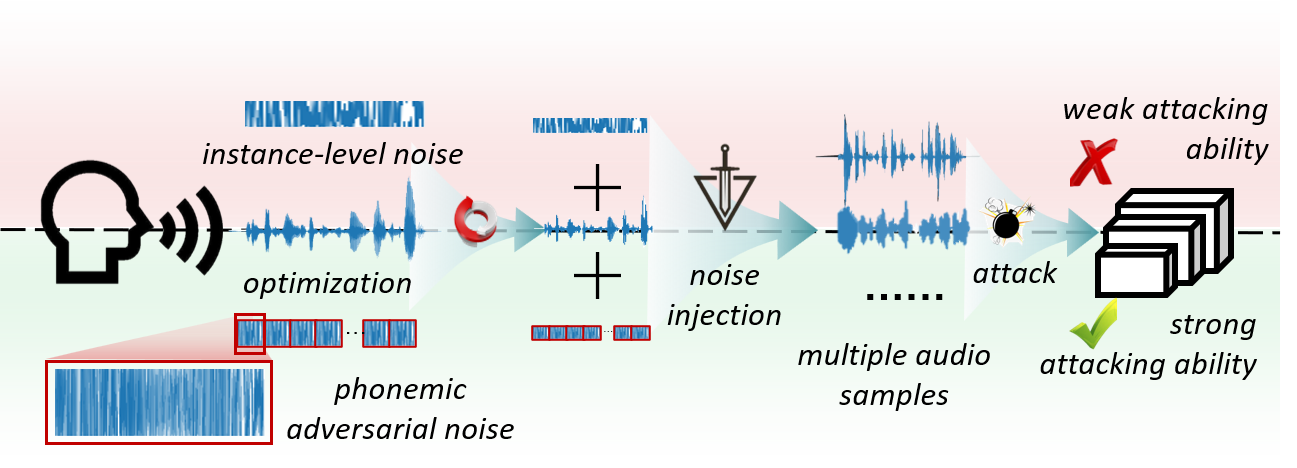}
\caption{Generating  phonemic adversarial noise to perform better attacks for audio recognition. Our phonemic noise enjoys faster generation speed and stronger universal attacking ability compared with the existing coarse-grained adversarial noises in instance level.}
\label{fig:waveform}
\end{figure}

Up to now, there exists a series of work to explore such adversarial attacking strategies for audio recognition task \cite{li2020universal,lu2021exploring}. Generally, a line of the previous work aims to specially generate adversarial noises for each audio. This kind of adversarial examples always has strong attacking ability due to the well fine-training to certain data points. Also, these generated adversarial audio noises could mislead models to output particular results (\ie, target attack) or random results (\ie, non-target attack). Another line of the existing work aims to produce adversarial audio noises which could make effectiveness to arbitrary input audio, \ie, universal adversarial noise. This kind of adversarial examples are often non-target attacks since the influenced audio data is unpredictable.

Though a variety of attempts have been tried for audio adversarial noise generation, existing methods always concentrate on a coarse understanding, to wit, ignoring the fundamental phonemic characteristic of audio speech, making them far away from satisfactory. In a nutshell, the existing adversarial attacks show weak attacking ability with such limitations that can be summarized as: (1) performing adversarial attacks is time expensive, that means, one has to cost much time in generating adversarial noises, making it not suitable enough in real-world scenarios; (2) there is still a distance from the satisfied universal attacking ability of current methods for diverse inputs, \ie, using one noise to attack multiple audio data.

To address the mentioned problem, we propose a fast \textbf{p}honemic \textbf{a}dversarial a\textbf{t}tack (PAT) paradigm with full consideration of the fundamental phonemic features to perform stronger audio attacks. Motivated by the observations that phoneme is the fundamental characteristic of audio speech according to linguists' studies \cite{10.2307/522070}, we aim to perform strong adversarial attacks against audio recognition models. It is reasonable that once we could optimize adversarial noises with full exploitation of critical phonemic features, the generated audio noises could achieve better attacking performance, \ie, fast and universal attacking ability. 
Specifically, regarding accelerating the generation speed, we acquire the representative instances (\ie, quantity less but phonemic features abundant) with the proposed phoneme density balanced sampling. Therefore, the heavy training dependency could be efficiently alleviated, \ie, enjoying significant generation speed up with comparable attacking performance. 
As for promoting universal attacking ability, we optimize the phonemic noises asynchronously onto audio instances during optimization via the proposed sliding phonemic noise injection. Thus, the adversarial noise will undergo different audio pieces that contain diverse phonemic features, resulting in better activation to model uncertainty.

For comprehensively evaluating the proposed PAT method, we mainly employ several awesome models in ASR task, including DeepSpeech2. Besides, we also perform a black-box evaluation on the VPR task, to wit, generating adversarial noise on ASR and straightly evaluating on VPR. The thorough experiments not only demonstrate the superior attacking ability of the proposed method compared with other state-of-the-art adversarial noises, but also strongly support the perspective of critical phonemic features.
To sum up, our main contributions can be summarized as follows:\par
\begin{itemize}
    \item To the best of our knowledge, we are the first to introduce the fundamental phoneme concept into audio attacks and propose the PAT paradigm for strongly attacking against audio recognition models.
    \item We propose the phonemic adversarial attack framework, which consists of dynamic noise duration calculation, phoneme density balanced sampling, and siding phonemic noise injection, to fast generate strong adversarial noise with universal attacking ability.
    \item We conduct extensive experiments on multiple audio tasks in both digital and physical world, demonstrating that our proposed PAT method outperforms other SOTA methods by large margins, \ie, at least \textbf{11}$\times$ speed up and \textbf{78\%} attacking ability improvement on ASR task.
\end{itemize}

\section{Related Work}
\subsection{Audio Recognition in Real World}

In recent years, intelligent audio recognition has attracted much attention, promoting the flexibility of our daily life, including automatic speech recognition (ASR) and voiceprint recognition (VPR). 
ASR task mainly aims to directly transform speech waveforms or feature sequences into text using deep learning techniques \cite{zhang2018deep}. For example, DeepSpeech \cite{hannun2014deep} and its improved versions \cite{amodei2016deep} are typical in this sub-area. Besides, the DFCNNs \cite{zhang2018fault}, which consists of superimposed convolutional layers to store more historical information, also obtain higher recognition accuracy. Among the studies of the VPR task, typical methods such as x-vector \citep{snyder2018x}, which uses a time-delay network as a deep feature extraction network, and DeepSpeaker \citep{li2017deep}, which improves recognition accuracy using deep residual RNN and triplet loss, also show considerable performance.


\subsection{Adversarial Attacks for Audio}

Adversarial example, which misleads deep models into wrong predictions, has been proposed first in computer vision \cite{goodfellow2014explaining,feng2020adversarial,zhang2021attack,DBLP:conf/ijcai/Wang21,DBLP:journals/corr/abs-2204-06213,Wang_2021_CVPR,9632406}. In recent years, this kind of studies also attracts the attention of researchers in the audio area, such as \cite{carlini2018audio,qin2019imperceptible,liu2020weighted,xie2021enabling,neekhara2019universal,senior2020stop,xie2020real,zong2021targeted,yakura2018robust,zong2021targeted}.
Unfortunately, despite achieving some certain results, these studies still show weaknesses. For example, \cite{carlini2018audio,qin2019imperceptible} can make audio transcribe to any text set by the adversary through enough iterations, but the massive time cost make it infeasible in practice. \cite{liu2020weighted,xie2021enabling} alleviate the huge computation cost while lacking the universal attacking ability. Recently, \cite{zong2021targeted} proposes a universal target attack but reporting not satisfactory performance. In this paper, we aim to generate strong adversarial noises in the real world based on the in-depth understanding of fundamental audio characteristics, \ie, phonemes.

\section{Approach}

In this section, we will first introduce the problem definition and then give a thorough elaboration about the proposed \textbf{p}honemic \textbf{a}dversarial a\textbf{t}tack (PAT) method that could generate strong universal adversarial noises fast.

\subsection{Problem Definition}

Given an audio dataset $\mathcal{X}$, an adversarial audio noise $\delta$ could mislead a given deep model $\mathcal{F}(\cdot)$ into wrong predictions by perturbing a clean audio $x$ where $x \in \mathcal{X}$, and the noise $\delta$ is subject to a $\varepsilon$-constraint. Furthermore, for a universal adversarial audio noise, its fooling ability is much stronger, \ie, effective to almost all $x \in \mathcal{X}$,  which can be formulated as:
\begin{equation}
    \begin{aligned}
        \mathcal{F}(x+\delta)& \neq \mathcal{F}(x)\quad s.t. \quad \|\delta\|<\varepsilon, \\
        &\textit{for almost all} \quad x \in \mathcal{X}
    \end{aligned}
\end{equation}
where $\|\cdot\|$ is a distance metric which measures the noise distortion magnitude by $l_{\infty}$-norm, and $\varepsilon$ is a constraint value.

However, in this paper, we aim to propose the phonemic adversarial attack (PAT) paradigm, which is significantly different from the above definition. More precisely, since our PAT takes full advantage of the phonemic information, it can be formulated as follows:
\begin{equation}
\begin{aligned}
    \mathcal{F}(x+\mathcal{C}(\delta_{p}, n))& \neq \mathcal{F}(x)\quad s.t. \quad \|\delta_{p}\|<\varepsilon, \\
        &\textit{for almost all} \quad x \in \mathcal{X}
\end{aligned}
\end{equation}
where $\mathcal{C}(\delta_{p}, n)$ is a function that splices $n$ pieces of phonemic noises $\delta_{p}$. Specifically, $\mathcal{C}$ can be formulated as:
\begin{equation}
    \mathcal{C}(\delta_{p}, n) = \begin{matrix} \underbrace{ \delta_p\oplus \delta_p\oplus \cdots \oplus\delta_p } \\ n \end{matrix}.
\end{equation}

Under our PAT paradigm, the adversarial noise duration is at phoneme level, which makes it necessary to splice dozens of phonemic noises circularly, \eg, for an audio $x$ with duration $l_x$, we need to splice $n = \frac{l_x}{l_{\delta_p}}$ pieces of noises, where $l_{\delta_p}$ is the duration of phonemic noise $\delta_{p}$.

\subsection{Overview}
\begin{figure*}[t]
\begin{center}
\includegraphics[width=0.99\linewidth]{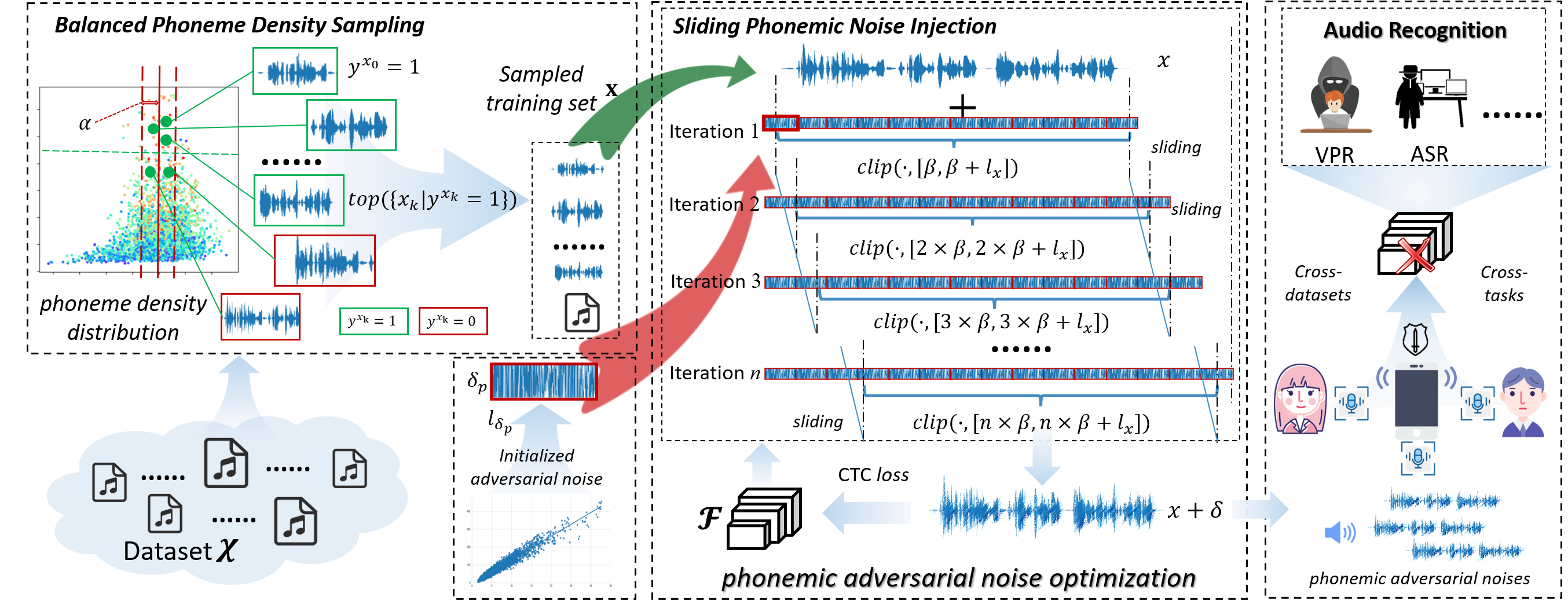}
\end{center}
\caption{The \textbf{p}honemic \textbf{a}dversarial a\textbf{t}tack (PAT) framework that adversarially mislead audio recognition models at phoneme level. This phonemic adversarial noise can be generated fast and achieves strong universal attacking ability in real world.}
\label{fig:framework}
\end{figure*}

According to the linguists' studies \cite{10.2307/522070}, we find such observation that phoneme plays a fundamental role in human speech. In other words, it could be regarded that different speeches are the different combinations of multiple phonemes. This observation encourages us to consider deeper exploitation of phonemic features inside audio instances, \ie, performing attacks with respect to phonemes. 
To this end, we proposed the \textbf{p}honemic \textbf{a}dversarial a\textbf{t}tack (PAT) paradigm to promote better adversarial attacks, \ie, faster and stronger universal attacking ability, by phoneme density balanced sampling and sliding phonemic noise injection, respectively. The framework of the proposed PAT is shown in Figure \ref{fig:framework}.

Regarding accelerating the generation speed of the adversarial noises, we mainly aim to alleviate the heavy training dependency, \ie, a large amount of training data, by the \textbf{phoneme density balanced sampling}. Since our PAT targets to attack phonemic features, it is possible for us to sample those typical instances with abundant phonemic patterns out for training. Therefore, we are allowed to utilize these limited instances to optimize the phonemic adversarial noises for achieving comparable attacking performance but cheaper generation time cost.
As for promoting universal attacking ability to various input audio, we aim to improve the phoneme diversity during noise optimization progress by \textbf{sliding phonemic noise injection}. Since the phonemic noises are injected into audio instances in an asynchronous way with the sliding window, it is easier for adversarial noises to capture the critical patterns among diverse phonemic features. Thus, the generated adversarial phonemic noises can be more pertinence against the fundamental phonemic features and activate model uncertainty for stronger attacking ability, \ie, universal.

\subsection{Phoneme Density Balanced Sampling}

In general, one of the reasons that the existing universal attacking studies spent too much time on adversarial noise generation is the heavy dependency on training data, \ie, a large amount of audio have to be employed. Unfortunately, previous work mainly optimize the adversarial noises at instance level, which makes them impossible to alleviate the straining scale. However, in our PAT paradigm, the phonemic-feature-oriented attacking framework allows optimizing the phonemic adversarial noises with fewer training instances. More precisely, since the phonemic features are extremely brief compared with the instance-level data (\ie, usually dozens of times shorter), one instance might contain dozens of different phonemic features, making it feasible to well optimize the phonemic noise with fewer instances. Based on this viewpoint, therefore, we sample the more representative instances from the given dataset under the consideration of the phoneme density, \ie, phoneme density balanced sampling.

Phoneme density, which is defined by us as the phoneme number in a unit duration, is the sampling basis for selecting the more representative instances. In other words, phoneme density depicts the enrichment of phonemes inside a certain data point. The phoneme density of audio $x$ can be formulated as $\frac{counter(\psi(x))}{l_x}$, where $\psi(\cdot)$ is a phoneme recognizer \cite{g2pE2019}. Furthermore, considering the correlations between instance and dataset, we deem that representative data points might have relevantly balanced phoneme density correlated to the average phoneme density of the dataset. That is, the phoneme density should be consistent with the overall phoneme distribution at the dataset level as far as possible. The insight behind this viewpoint is that too high or too low phoneme density might arise in over-fitting or under-fitting in practice.

Specifically,  for each sample $x_{k}\in \mathcal{X}$, we will estimate its phoneme density and decide if it should be a sampled instance, following the formulation below: 
\begin{equation}
y^{x_{k}} = 
\begin{cases}
1, \mbox{if \quad$\lvert\textit{D} - \frac{counter(\psi(x_k))}{l_{x_k}}\rvert \leq \alpha$}\\
0, \mbox{else}
\end{cases},\\
\label{eqn:trainingdata}
\end{equation}

where $\textit{D}$ is the average phoneme density of the given dataset which is calculated by $\textit{D} = \frac{\sum_i counter(\psi(x_i))}{\sum_i l_{x_i}}, x_i \in \mathcal{X}$, and ${\alpha}$ is a threshold that controls the density interval that if an instance is acceptable. Finally, we sample the training data points $\textbf{x} = top(\{x_k|y^{x_k} = 1\})$ following a simple picker function $top(\cdot)$ that select the toper samples with longer duration and amplitude, \ie, sampling 10, 5, even 1 instances. 

Since the sampled training instances possess relevantly rich phonemic features, they can be very efficient and targeted training samples for phonemic adversarial noises. Besides, compared with the normal optimization process, the scale of the sampled audio data is much lower, bringing additional computation cost decreasing.
\subsection{Sliding Phonemic Noise Injection}

In our phonemic adversarial attack paradigm, the most critical characteristic that decides the attacking ability is the phoneme-oriented influence from the generated noises. Considering the previous observations that feature diversity might be critical for model generalization \cite{DBLP:journals/jbd/ShortenK19,DBLP:conf/acl/FengGWCVMH21}, we are motivated to improve the diversity of the employed phonemic features during optimization for better attacking ability. Therefore, we propose the sliding phonemic noise injection that forces the phonemic adversarial noises to asynchronously undergo different audio pieces, instead of simply matching the sampled audio aligned, therefore capturing critical fundamental patterns among diverse phonemic features. 
To be specific, different from the previous adversarial noise optimization process that injects the noise $\delta$ to be optimized aligned into an audio instance $x$, our sliding phonemic noise injection specially introduces a sliding window. In this way, for each iteration, the phonemic noise will be injected asynchronously onto different audio pieces with a sliding step $\beta$. For example, for a phonemic noise $\delta_p$, if it is injected onto the audio piece $[m, n]$ in the first iteration. Then for next iteration, it will be slid $1\times\beta$ duration and undergo the piece with duration interval $[m+\beta, n+\beta]$. To sum up, the sliding window helps the phonemic noise facing diverse audio pieces, therefore leading to enhancement of the phonemic diversity in disguise.

In practice, during sliding phonemic noise injection process, given a phonemic noise $\delta_p$, an instance $x$, and the sliding step $\beta$, we can inject the phonemic noise by the following injection module $\mathcal{A}$:
\begin{equation}
     \mathcal{A}_{n}(x, \delta_p, \beta)=clip(\mathcal{C}(\delta_p, \lfloor\frac{l_{x}}{l_{\delta_p}}+n\rfloor), [n\times\beta, n\times\beta+l_{x}]) 
\end{equation}
where $clip(\cdot, [start, end])$ is a clipping function that clips an audio into a designated interval $[start, end]$, $l_{x}$ is the length of the sampled instance $x$ during training, $l_\delta$ is the length of the phonemic noise $\delta_p$, $\mathcal{C}(\cdot, \cdot)$ is a splice function, $\beta$ is the step length of the sliding window, and $n$ is the iteration rounds.

Benefiting from the sliding phonemic noise injection, the phonemic noise to be optimized will undergo diverse phonemic features inside the sampled instances in each iteration. In this way, the adversarial noise to be optimized could capture the critical phonemic patterns much more efficiently, resulting in better activation to model uncertainty, \ie, stronger attacking ability. 


\subsection{Overall Optimization}
In our phonemic adversarial attack paradigm, we perform strong adversarial attacks (\ie, fast and universal) according to the in-depth exploitation of phonemic features among audio speeches. For accelerating the generation speed, we first acquire the representative instances by the proposed phoneme density balanced sampling. For promoting the universal attacking ability, we enhance the phonemic features diversity by conducting the sliding phonemic noise injection strategy. In this way, the generated phonemic adversarial noises will take full advantage of the phoneme-level optimization, therefore achieving better adversarial attacking performance.

Besides the above elaborations, we are also concerned about the real effectiveness in the complex physical world environment. Considering that previous studies have made some efforts to this end, we just introduce the room impulse response (RIR) \cite{peddinti2015reverberation}, which is one of the most effective approaches by encoding the acoustic channel state information (CSI) for guaranteeing the usability in the real world.

Specifically, given an audio dataset $\mathcal{X}$ and target model (\eg, an ASR model) $\mathcal{F}$, the phonemic adversarial noise generation progress for audio recognition can be formulated as:
\begin{equation}
\label{eqn:finalfunction}
    \begin{aligned}
    \min-\mathcal{L}(\mathcal{F}&(\mathcal{R}(x+\mathcal{A}_{n}(x, \delta_p, \beta)),\mathcal{F}(x)), \\
&\emph{s.t.}\quad \lVert\delta_p\rVert_{\infty}\leq\varepsilon \quad x\in\textbf{x} 
\end{aligned}
\end{equation}
where $\mathcal{R}(\cdot)$ is the RIR function, $\lVert\cdot\rVert_{\infty}$ is a $l$-$\infty$ norm. The overall training progress can be found in Algorithm \ref{alg:algorithm}.

\begin{algorithm}[tb]
\caption{Phonemic Adversarial Attack}
\label{alg:algorithm}

\textbf{Input}: target model $\mathcal{F}$, audio dataset $\mathcal{X}$, allowed perturbation constraint $\varepsilon$, the iteration number $\mathbf{iters}$, the epoch number $E$, phoneme recognizer $\psi(\cdot)$, and RIR module $\mathcal{R(\cdot)}$, $\alpha$, and $\beta$\\
\textbf{Output}: phonemic noise $\delta_p$ 
\begin{algorithmic}[1] 
\STATE \# sample the training data
\STATE select the $\textbf{x}$ according to the result of Eqn (\ref{eqn:trainingdata})
\STATE initialize the $\delta_p$ randomly
\FOR{i\;=\;1\;up\;to\;$E$}
\FORALL{$x \in \textbf{x}$}
\STATE $n = 1$
\WHILE{$n\;<=\;\operatorname{iters}\;$}
\STATE calculate $\mathcal{A}_{n}(x, \delta_p, \beta)$
\STATE optimize the $\delta_p$ with Eqn (\ref{eqn:finalfunction})
\ENDWHILE
\ENDFOR
\ENDFOR
\STATE \textbf{return} $\delta_p$
\end{algorithmic}
\end{algorithm}

\section{Experiments}
In this section, we first introduce the detailed experimental settings, and then we evaluate the proposed adversarial noise generation framework from multiple perspectives and report the experimental results. Further, we also provide some ablations and discussions to better understand the adversarial noise in audio area.
\subsection{Experimental Settings}
\subsubsection{Dataset and Models}
Regarding the dataset, we mainly employ the well-known LibriSpeech dataset \cite{panayotov2015librispeech} for ASR and VPR.  For validating the universality of the generated adversarial noises, we also employ the TED-LIUM dataset \cite{rousseau2012ted} and Mozilla Common Voice Dataset \cite{ardila2019common}. As for the compared models, we pick multiple state-of-the-art models with different architectures and parameter scales for different tasks. Specifically, for ASR task, we select DeepSpeech2, Wav2Vec, and SEW models. For VPR task, we select DeepSpeaker models. Note that all of these models are pre-trained on the open-source dataset.

\subsubsection{Evaluation Metrics}
To evaluate the performance accurately, we introduce several widely-used metrics for different tasks. For ASR task, we employ the character error rate (CER) and attacking success rate (SR) as the evaluation metrics. Also, the $dB_{x}(\delta)$ is introduced for constraining the noise intensity.
For VPR task, we simply introduce the \emph{Accuracy}, which is also the widely used evaluation metric in this sub-area, as the evaluation basis.

\subsubsection{Compared Methods and Implementation details}

We choose several state-of-the-art works in adversarial audio noise generation, especially universal adversarial audio noise methods. Specifically, we select the C\&W attack (``C\&W'') \cite{carlini2018audio}, Targeted Universal Adversarial Perturbations (``TUAP'') \cite{zong2021targeted}, and Crafting Audio-Based UAP (``CUAP'') \cite{senior2020stop} as compared baselines. Besides, we also introduce some typical noises from Noise92 \cite{noise92url} for comparison.  The implementation details can be found in \emph{supplementary materials}.

\subsection{Attacking Ability of PAT}

For demonstrating the effectiveness of the proposed PAT framework, we first conduct experiments in the both digital and physical worlds.


As for digital world attack, we train our adversarial noise with the selected data points sampled from the LibriSpeech dataset
Besides the baselines, we also introduce some simulated noises from Noise92 as a comparison. The experimental results can be witnessed in Table \ref{tab:digital} (\emph{more results in supplementary materials}). It is distinct that our generated adversarial noises achieve better performance, \ie, attacking ability. To conclude, there are several meaningful insights:
\begin{itemize}
    \item Our phonemic adversarial attack enjoys strong universal attacking ability due to its special design. In detail, the attacking success rate (SR) of ``Ours'' has a great improvement, \ie, 0.87, and the character error rate (CER) of ``Ours'' also achieves a very high value, \ie, 0.83. Compared  with the baseline methods, our PAT noise show \textbf{66.67\%} improvement on average for SR and \textbf{51.67\%} improvement on average for CER, showing strong universal attacking capability.
    \item The proposed PAT is able to generate universal adversarial audio noises with lower time cost, \ie, much faster than the comparisons. For example, the typical ``C\&W'' attack could only achieve 0.17 CER, but its cost is 156 mins (\textbf{10+ times} more than ``Ours''). The SOTA universal attacking ``CUAP'' achieves strong attacking ability but spends more computation cost, \ie, even up to \textbf{40+ times} additional expenses. 
    \item The C\&W attack shows a very low attacking success rate and character error rate compared with other methods. We attribute to the basic that C\&W is not special-designed for universal attacking. Besides, it should be noted that C\&W has higher $\varepsilon$ because the constraint of C\&W is not so strict, which further sets our phonemic adversarial noise off to advantage. 
\end{itemize}

Besides, we employ a LibriSpeech-pre-trained DeepSpeaker model, a typical and well-used architecture for VPR, to validate the cross-task attacking ability. According to the experimental results, as shown in Table \ref{tab:vprresults}, we can conclude that our PAT could attack cross tasks, demonstrating its strong universal attacking ability.

\begin{table}[h]
\centering
\begin{center}   
\caption{Experimental results on the LibriSpeech dataset.}  
\label{tab:digital} 
\begin{tabular}{cp{2cm}<{\centering}cp{1cm}<{\centering}cp{2cm}<{\centering}cp{1cm}<{\centering}cp{1cm}<{\centering}} 
\toprule
Method & Time(mins) & $dB_{x}(\delta)\textcolor{green}{\downarrow}$ & SR$\textcolor{green}{\uparrow}$ & CER$\textcolor{green}{\uparrow}$ \\   
\midrule
	Raw & - & - & - & 0.07 \\ 
\midrule
    Noise & - & -33 & 0.01 & 0.07 \\ 
\midrule
	C\&W & 156 & -20 & 0.06 & 0.17 \\ 
\midrule
	TUAP & 187 & -33 & 0.06 & 0.18 \\ 
\midrule
	CUAP & 575 & -33 & 0.49 & 0.59 \\ 
\midrule
	Ours & 14 & -33 & \textbf{0.87} & \textbf{0.83} \\ 
\bottomrule
\end{tabular}   
\end{center}   
\end{table}

\begin{table}[]
    \centering
    \caption{The experimental results on VPR task. Note that we adopt a full black-box setting.}
    \begin{tabular}{ccccccc}
    \toprule
    Method & Benign & C\&W & CUAP & TUAP & Ours \\\midrule
    Accuracy$\textcolor{green}{\downarrow}$ & 87.5 &  62.5 & 30.0 & 30.0 & \textbf{21.0} \\\bottomrule
    \end{tabular}
    \label{tab:vprresults}
\end{table}


As for the physical world attack, we construct a simulated scenario in the physical world and report the experimental results.
Specifically, we employ 3 devices (\ie, 2 iPads and 1 cellphone) as main broadcaster and receptor. One of the 2 iPads is employed for playing noise audio, another is employed for playing benign audio, and the cellphone is employed to receive the physical-mixed audio. We provide more details and examples in the \emph{supplementary materials}.
For evaluation, we randomly select 50 instances from the LibriSpeech and play them for testing, and the experimental results are shown in
Table \ref{tab:phyresults}, it can be witnessed that compared with the baselines, our PAT achieves the best attacking ability, \ie, the SR and CER of Ours are respectively \textbf{0.68} and \textbf{0.74}, which is higher than those of CUAP, \ie, 0.32 and 0.48. Besides, referring to Table \ref{tab:digital}, the digital-physical drop of ours is less than that of other baselines, showing the fundamental phonemic features is of great significance for promoting adversarial attacks.


\begin{table}[]
\caption{Experimental results in the physical world.}
\centering
\begin{tabular}{clllll}
\toprule
Method & Noise & C\&W & TUAP & CUAP & Ours \\
\midrule
SR$\textcolor{green}{\uparrow}$     &  0.02 & 0.04 &  0.04 & 0.32 & \textbf{0.68}    \\\midrule
CER$\textcolor{green}{\uparrow}$   &  0.11 &  0.22 &  0.21 & 0.48 & \textbf{0.74}    \\\bottomrule
\end{tabular}
\label{tab:phyresults} 
\end{table}

\subsection{Ablation Study}

Since there are two key operations in our proposed PAT framework, it is necessary to perform detailed ablations on them for investigating their effectivenesses further. Besides, we also explore the sound effects of the hyperparameters, such as $l_{\delta_{p}}$, $\alpha$, and $\beta$.

\subsubsection{The Effectiveness of Key Operations}

For investigating the effectiveness of the key operations inside our proposed PAT, we design an ablation that controls their employment, \ie, phoneme density balanced sampling (``PDBS'') and sliding phonemic noise injection (``SPNI'').

Specifically, when adopting the controlled key operation, we would not adopt the other operations, \eg, when adopting the ``PDBS'', the ``SPNI'' will not be employed. To make the results more convincing, we provide multiple results with diverse experimental settings as comparisons. The experimental results can be witnessed in Table \ref{tab:ablationkey}. We can naturally draw such conclusions: 
1) the ``PDBS'' significantly decreases the optimization cost during training, \ie, the training Time decrease from 557 min to 14 min;
2) the ``SPNI'' show greater effectiveness on improving the universal attacking ability, \ie, compared with the normal settings (no PDBS and SPNI), the SR and CER of ``SPNI'' increase respectively from 0.14, 0.27 into 0.40, 0.52. Besides, we have to notice that the performance of the whole PAT is much higher compared with that of the ``SPNI'' only, we conjecture the insight behind this observation is that the possible overfitting to phonemic features is evaded, therefore allowing the utility of ``SPNI'' could be maximized. 
\begin{table}[!]   
\begin{center}   
\caption{Ablations on the key factors of ours framework}  
\label{tab:ablationkey} 
\begin{tabular}{ccccc}   
\toprule
PDBS & SPNI & Time (min) & SR$\textcolor{green}{\uparrow}$ & CER$\textcolor{green}{\uparrow}$ \\
\midrule
 $\times$ & $\times$  & 557 & 0.14 & 0.27 \\
\midrule
 $\checkmark$ & $\times$ & 14 & 0.39 & 0.50 \\
\midrule
$\times$ & $\checkmark$ & 557 & 0.40 & 0.52 \\
\midrule
$\checkmark$ & $\checkmark$ & 14 & \textbf{0.87} & \textbf{0.83} \\
\bottomrule
\end{tabular}   
\end{center}   
\end{table}
\subsubsection{The Effectiveness of Hyperparameters}
In our proposed PAT framework, there exists several hyperparameters, \eg, $l_{\delta_p}$, $\alpha$, and $\beta$. It is necessary to make it clear that how these hyperparameters impact the framework performance. Thus, we conduct several ablations on them and report the experimental results for further understanding of PAT.

\begin{figure}
    \centering
    \subfigure[ $l_{\delta_{p}}$]{
    \includegraphics[width=0.3\linewidth]{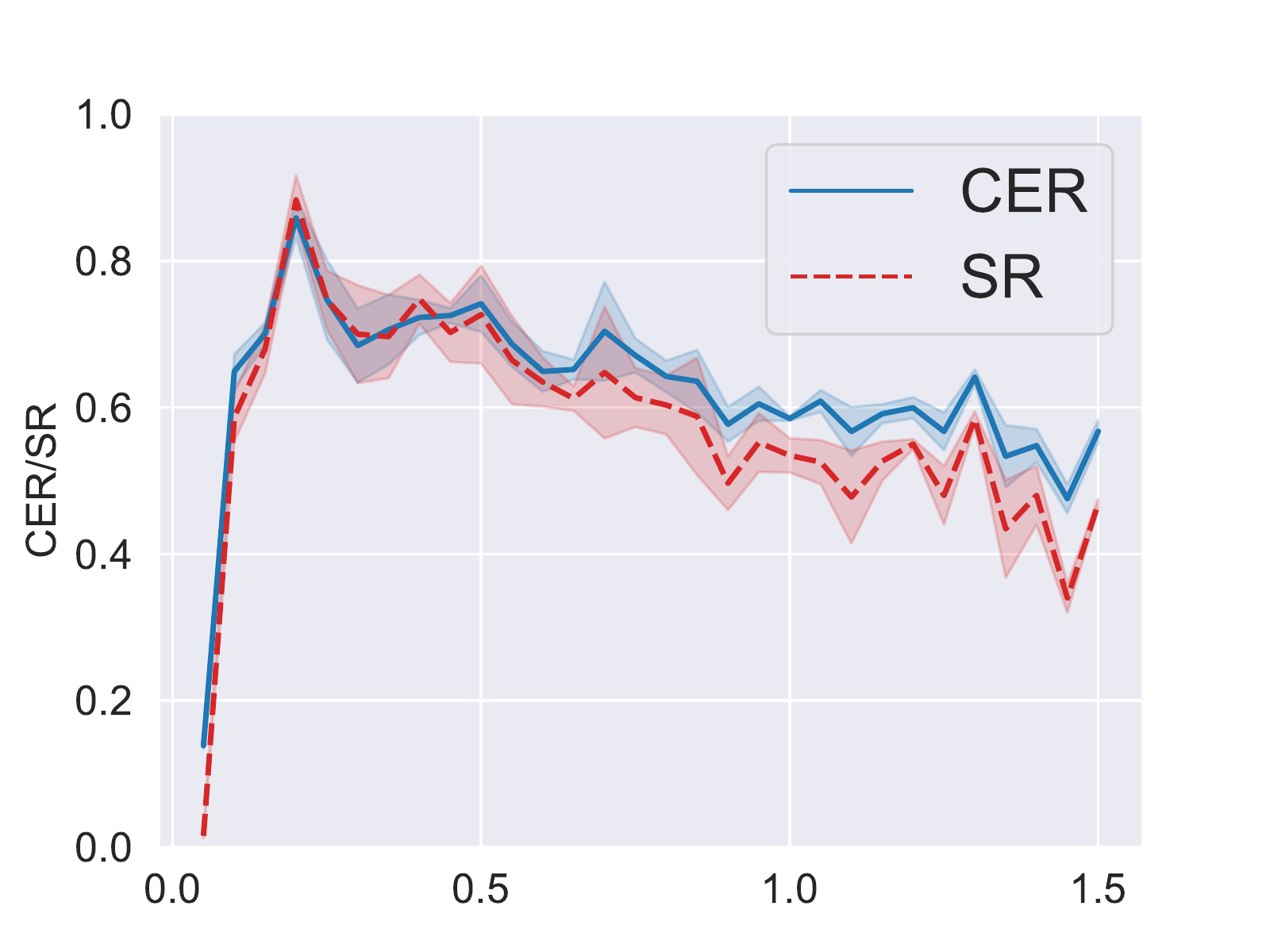}
    \label{fig:ablation-l}}
    \subfigure[$\alpha$]{
    \includegraphics[width=0.3\linewidth]{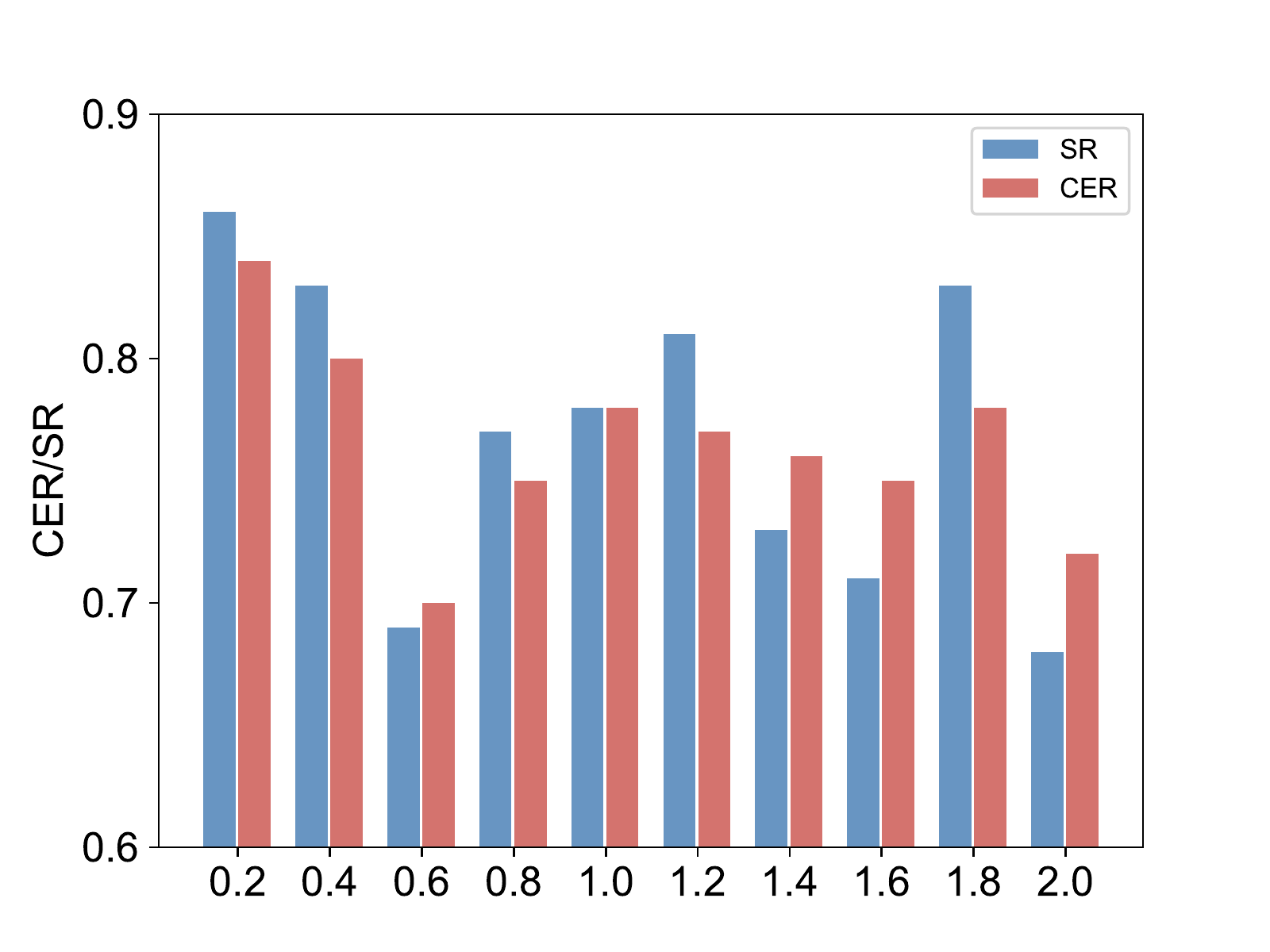}
    \label{fig:ablation-alpha}}
    \subfigure[$\beta$]{
    \includegraphics[width=0.3\linewidth]{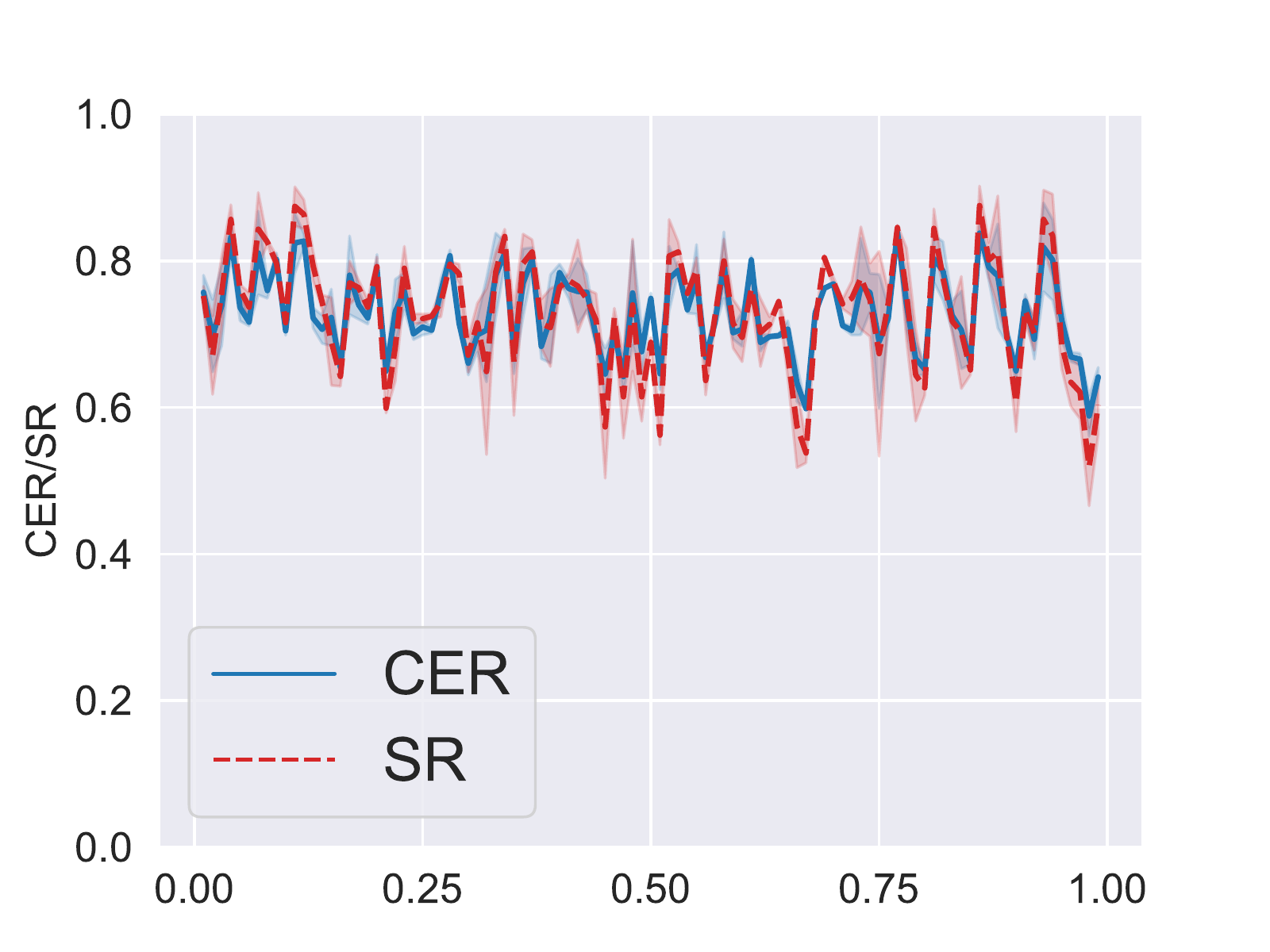}
    \label{fig:ablation-beta}}
    \caption{The sub-figure (a), (b), and (c) are respectively the ablation results on the hyperparameter $l_{\delta_p}$, $\alpha$, and $\beta$. Note that the experiments are conducted on LibriSpeech dataset.}
\end{figure}

Specifically, as for the $l_{\delta_p}$, we deem that it is highly correlated to the phoneme duration. Thus, we constrain its sampling range as (0, 1.5] and set the interval as 0.05 (\ie, 30 sampled points in total). Then we optimize the adversarial audio noise and test its attacking ability.
Similarly, as for $\alpha$ and $\beta$, we also adopt an interval sampling strategy, but the sampling ranges of $\alpha$ and $\beta$ are respectively (0, 2] (interval is 0.2, 10 sampled points in total) and (0, 1) (interval is 0.01, 99 sampled points in total).
The results can be found in Figure \ref{fig:ablation-l}, \ref{fig:ablation-alpha}, and \ref{fig:ablation-beta}.
According to the results, as for $l_{\delta_p}$, we can find that the performance of the generated adversarial noises achieves a relevantly higher level when $l_{\delta_p}$ is 0.20. We conjecture that this duration makes the noise learn the adversarial phonemic features against the audio more efficiently, \ie, achieving the balance point of the trade-off between over-fitting and under-fitting. As for $\alpha$ and $\beta$, we find it has no explicit rule, \ie, keeping relevantly stable performance, we therefore set them as 0.2 and 0.77 empirically. It should be noted that for $\alpha$, the higher performance occurs with lower values. Besides, although the $\beta$ shows relatively instability, its worst case is also much better than the baselines, \ie, CER is 0.6 which is also higher than that of the better comparison CUAP, to wit, 0.59.
 
\subsection{Discussions and Analysis}
In this section, we further give some more insights about our proposed fast phonemic universal adversarial attack framework. Foremost, we try to answer a question that what makes phonemic adversarial noises work compared with other common adversarial noises. Further, we investigate the correlations between phoneme density and audio length, leading to better understanding about their effectiveness to noise training. Finally, we conduct simple black-box experiments for exploring the transferability of the proposed PAT.
\subsubsection{Why Phonemic Noises?}
To study the critical reason, what makes the phonemic noises perform well, behind the framework, we thus anticipate to analyse the diverse adversarial noises under the perspective of audio spectrogram.

\begin{figure}[b]
	\centering
	\includegraphics[width=0.99\linewidth]{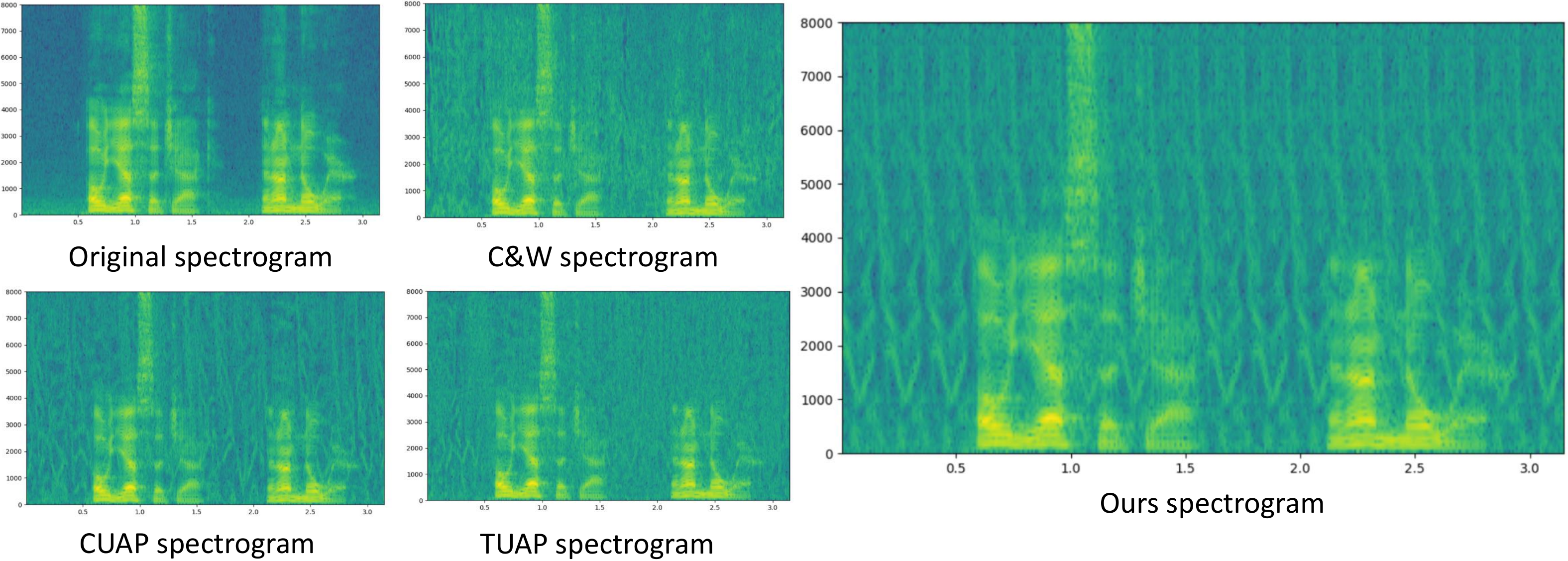}
	\caption{Spectrogram of different adversarial audio.}
	\label{fig:spectrogram}
\end{figure}

In detail, we employ the generated adversarial audio examples of ``C\&W'', ``TUAP'', ``CUAP'' and ``Ours'', then draw their spectrogram as shown in Fig \ref{fig:spectrogram}. Besides, we further provide more spectrograms of our PAT with different audio noise lengths as longitudinal comparisons (\emph{more results can be found in supplementary materials}). In general, we can draw some meaningful qualitative conclusions as: 1) for the common adversarial attacks, like ``C\&W'' \etc, they perturb the audio data into undistinguished patterns in spectrogram view, therefore making the recognizable patterns in this level more difficult for models to capture; 2) our PAT, however, not only perturb the spectrogram but also make the spectrogram show strong cyclical patterns, which further mislead the models to capture wrong patterns. To sum up, our proposed method attacks the audio models by  both confusing the right spectrogram patterns and injecting wrong spectrogram patterns, resulting in stronger attacking ability.

\subsubsection{The Cross-dataset Generalization}
Compared with universal attacking ability on one dataset, evaluating the adversarial noises' generalization on diverse dataset make it much more convincing. Thus, we perform additional experiments on investigating the attacking capacity when generating based on dataset, but testing based another. We employ LibriSpeech ($\mathcal{X}_1$), TED-LIUM ($\mathcal{X}_2$), and Mozilla Common Voice Dataset ($\mathcal{X}_3$).
We use $\mathcal{X}_1\rightarrow\mathcal{X}_2$ to denote that the adversarial noises are generated on $\mathcal{X}_1$ and tested on $\mathcal{X2}$. According to Table \ref{tab:crossdataset}, we could conclude that our proposed PAT achieves stronger attacking ability in most cases, especially in cross-dataset evaluations, \ie, in setting $\mathcal{X}_1\rightarrow\mathcal{X}_2$, the CER is \textbf{0.43}, where that of C\&W, CUAP, and TUAP are 0.26, 0.36, and 0.25, strongly demonstrating the universality of PAT. \emph{See supplementary materials for more details.}

\begin{table}[!]
\caption{Cross-dataset generalization validation results.}
\label{tab:crossdataset}
\begin{tabular}{cccccc}
\toprule
Setting & Metric & C\&W & CUAP & TUAP & Ours \\\midrule
\multirow{2}{*}{$\mathcal{X}_1\rightarrow\mathcal{X}_2$} & SR$\textcolor{green}{\uparrow}$    &      0.08 &  0.20   &   0.07   &   \textbf{0.33}   \\ \cline{2-6} \noalign{\smallskip}
& CER$\textcolor{green}{\uparrow}$   &  0.26    & 0.36     &   0.25   &    \textbf{0.43}  \\\midrule
\multirow{2}{*}{$\mathcal{X}_1\rightarrow\mathcal{X}_3$} & SR$\textcolor{green}{\uparrow}$    &     0.32 &  0.66    &    0.36  &  \textbf{0.70}    \\\cline{2-6} \noalign{\smallskip}
& CER$\textcolor{green}{\uparrow}$   &    0.41  &  \textbf{0.67}    &  0.42    &   \textbf{0.67}
                      \\\bottomrule
\end{tabular}
\end{table}

\subsubsection{The Transferability of the Phonemic Adversarial Noise}
Since our PAT attacks fundamental phonemic features, we believe it might possess comparable transferability. To verify, we employ additional models in fashion, \ie, Wav2Vec ($\mathcal{F}_2$), and SEW ($\mathcal{F}_3$) (note that DeepSpeech2 is denoted as $\mathcal{F}_1$), to evaluate the black-box transferability of our PAT framework. We optimize the adversarial noise on $\mathcal{F}_1$ and test it on the others.
The results shown in Table \ref{tab:trans} indicates that the proposed method achieves stronger transferable attacking ability than the comparisons (\emph{more results can be found in supplementary materials}). Although the models have different architectures and parameters, their performance is also impacted by our generated adversarial noises, showing that the critical phonemic features are beneficial to perform adversarial attacks.

\begin{table}[!]
\centering
\caption{The experimental results of transferable attacking.}
\begin{tabular}{ccccccc}
\toprule
Models & \multicolumn{2}{c}{$\mathcal{F}_1$} & \multicolumn{2}{c}{$\mathcal{F}_2$} & \multicolumn{2}{c}{$\mathcal{F}_3$} \\ \midrule
Metric &  SR$\textcolor{green}{\uparrow}$  & CER$\textcolor{green}{\uparrow}$ &  SR$\textcolor{green}{\uparrow}$  & CER$\textcolor{green}{\uparrow}$    & SR$\textcolor{green}{\uparrow}$    & CER$\textcolor{green}{\uparrow}$      \\\midrule
C\&W &0.06  &0.17 &0  &0.05  &0  &0.10   \\ \midrule
TUAP & 0.06 &0.18 &0  &0.04 &0.01  &0.09  \\ \midrule
CUAP & 0.49 &0.59 &0.02  &0.09  &0.11  &0.25  \\ \midrule
Ours & \textbf{0.87} & \textbf{0.83} & \textbf{0.09} & \textbf{0.19} & \textbf{0.31} & \textbf{0.44} \\ \bottomrule
\label{tab:trans}
\end{tabular}
\end{table}

\section{Conclusion}

In this paper, we propose a phonemic adversarial attack (PAT) paradigm, which exploits the fundamental characteristic, \ie, phonemic features, among audio data to perform stronger adversarial attacks (\ie, fast and universal). 
Specifically, we propose the phoneme density balanced sampling to acquire fewer but representative instances for the phonemic adversarial noise optimization, therefore accelerating the generation speed. Moreover, we elaborate a sliding phonemic noise injection strategy to enhance the phonemic feature diversity with an asynchronous sliding window during optimization process, which is beneficial for promoting the universal attacking ability of the adversarial noises. Since taking advantages of the phoneme-level features, the generated phonemic adversarial noise could enjoy better performance in the real world.
To evaluate the effectivenesses of the proposed method, we conduct extensive experiments, containing thorough ablations and discussions. The results demonstrate that our PAT method outperforms other SOTA methods by large margins, \ie, at least 11$\times$ generation speed up and 78\% attack success rate improvement.

In this work, we perform success attacks for audio recognition and reveal the importance of the phoneme-related features. We believe this typical phonemic features can be exploited deeper and promote better research in the future.
In the future, we are also interested in enhancing audio models with the help of phonemes.

\bibliography{aaai22}

\end{document}